\documentclass[twocolumn,showpacs]{revtex4}

\usepackage{amsbsy}
\usepackage{graphicx}

\begin{document}

\title{Nonanalytical equation of state of the hard sphere fluid}

\author{Ji\v{r}\'{\i} Kolafa}
\email{jiri.kolafa@vscht.cz}
\homepage{http://www.vscht.cz/fch/en/people/Jiri.Kolafa.html}
\affiliation{Department of Physical Chemistry, Institute of Chemical Technology, 166 28 Prague 6, Czech Republic}
\date{\today}

\begin{abstract}
An equation of state of the hard sphere fluid which is not analytical
at the freezing density is proposed and tested.  The nonanalytical
term is based on the the classical nucleation theory and is able to
capture the observed ``anomalous increase'' of pressure at high
densities.  It is combined with the virial expansion at low densities.
\end{abstract}


\maketitle

\section{Introduction}

It is well known that liquids can be easily superheated or
supercooled \cite{Debenedetti}. Equations of state (EOS) describing
liquids and gases are constructed as analytical functions for apparent
practical and also theoretical reasons (thermodynamic approximations
usually lead to classical mean-field type theories).  These functions
continuously extend to metastable and in most cases also unphysical
unstable regions.  Phase equilibria are then calculated from these
functions.  Yet there are arguments casting doubt on this picture. 

The first argument is based on the nucleation process, see Review
\cite{Binder} and references therein.  Pure liquid (any state in
general) in a locally stable but globally metastable state undergoes
after a certain period of time a spontaneous transition to a more
stable state.  In three-dimensional systems, liquid/gas and
liquid/crystal transitions are of the first order and the process is
controlled by homogeneous nucleation: When a nucleus (droplet) of the
more stable phase grows above a certain critical limit, it keeps
growing until bulk liquid freezes or evaporates.  Since thermodynamic
quantities are time averages over functions of configurations, we
cannot calculate them with arbitrary precision.  It does not help to
change the size of the system because in a larger system the
probability of the critical fluctuation is larger, the survival time
of the metastable state shorter, just to compensate higher precision
of averages in the larger system.  Therefore a value of a
thermodynamic quantity in a metastable state is not a function at all
because it is a subject of ``essential error''.  This error may be
viewed as a statistical-mechanical analogue of the Heisenberg
uncertainty principle.

The second argument is based on the Ising ferromagnet which serves as
the simplest model of the first-order phase transition (in the
equivalent form of the lattice gas as a model of liquid-vapor
equilibrium).  It has been proven \cite{Ising} that the
magnetization below the Curie temperature is not an analytical
function of the magnetic field at the phase transition point and
therefore the magnetization cannot be analytically continued to the
metastable state.  (A function is analytical at a point if its Taylor
expansion about this point exists and converges to given function in a
neighborhood of the point.)  Approximate approaches to this ``essential
nonanalyticity'' problem are based again on the droplet model
\cite{Binder}.

The third hint comes from our experience with the hard sphere fluid.
Not only we and other authors \cite{Speedy} encountered the
impossibility of determining accurately the equation of state at large
densities (at deeply metastable region), but we also found that the
data cannot be easily fitted to common continuous functions.  There is
an ``anomalous'' increase of pressure at high densities.  In
contrast, the second derivative of the compressibility factor reaches
a maximum at high density.


These arguments led us to an attempt to use formulas derived from the
simplest version of the classical nucleation theory in combination
with the virial expansion to develop an equation of state which is not
analytical at the freezing point.

\section{Theory}

\subsection{Classical nucleation theory}

In the classical nucleation theory \cite{Debenedetti} the Gibbs energy
of a spherical droplet of solid phase of radius $r$ emerged in fluid
is estimated by
\begin{equation}
  \Delta G(r) = {4\over 3}\pi r^3\rho(\mu_{\rm s}-\mu_{\rm f}) + 4\pi r^2 \gamma,
\label{Gr}
\end{equation}
where $\mu_{\rm s}$ and $\mu_{\rm f}$ are the chemical potentials of
the solid and fluid phases, respectively, $\gamma$ is the s/f
interfacial energy (surface tension), $\rho$ is the reduced number
density (sphere diameter is unity).

For metastable fluid ($\rho>\rho_{\rm fp}$, where $\rho_{\rm fp}$ is
the fluid density at the freezing point) it holds 
$\mu_{\rm s} < \mu_{\rm f}$ 
and the Gibbs function $G(r)$ exhibits a maximum at $r=r^*$,
$$
  r^*
    = -{2\gamma\over\rho(\mu_{\rm s}-\mu_{\rm f})} 
    \approx {2\gamma\over (\rho-\rho_{\rm fp})A} .
$$
where we linearized the difference, $(\mu_{\rm s}-\mu_{\rm
  f})\rho\approx -(\rho-\rho_{\rm fp})A$, $A>0$.
As soon as the nucleus happens to reach radius $r^*$, it keeps growing.
The probability of this spontaneous freezing (per unit time and
particle) is proportional to
\begin{equation}
   \exp\left[-\Delta G(r^*)\over kT\right] 
 = \exp \left[-{\rm const} \over(\rho-\rho_{\rm fp})^2\right],
\label{prec}
\end{equation}
where const is a positive constant.
This is proportional to the ``metastable uncertainty'', the inherent
inaccuracy of measurements on the metastable state.

\subsection{Anomalous behavior in the stable region}

The stable phase ($\rho<\rho_{\rm fp}$) can be viewed as a fluid phase with
``virtual'' nuclei of solid phase of radius $r$.  The probability of
finding a nucleus of radius $r$ is
\begin{equation}
  {\rm Prob}(r) \propto \exp\left[-\Delta G(r)\over kT\right]  .
\end{equation}
Quantities as the compressibility factor, $Z=p/(\rho kT)$ ($p$ denotes
pressure, $T$ temperature, and $k$ the Boltzmann constant), are
sensitive to the volume of the nucleus, ${4\over3}\pi r^3$.  The
anomalous part (caused by ``fluctuating nuclei'') is obtained by
integrating over all sizes of the nuclei,
\begin{equation}
Z_{\rm anom} = {\rm const}'\times  \int_0^{\infty} r^3 \exp\left[-\Delta
  G(r)\over kT\right] {\rm d} r .
 \label{ZanomG}
\end{equation}
Substitution $r=x/(8\pi\gamma)^{1/2}$ and rearrangement of constants leads
to the normalized form
\begin{equation}
  Z_{\rm anom}(\rho) = \beta\, \psi_3(\alpha(\rho-\rho_{\rm fp})) ,
  \label{Zanom}
\end{equation}
where $\alpha$, $\alpha>0$, and $\beta$ are constants and
\begin{equation}
  \psi_n(t) = \int_{0}^{\infty} x^n
  \exp\left({x^3\over3}t-{x^2\over2}\right){\rm d} x .
\label{psin0}
\end{equation}
This integral converges for all $t\le 0$ (then  $\rho<\rho_{\rm fp}$
in Eq.~\ref{Zanom}), but function $\psi_n(t)$ is not
analytical at $t=0$ and cannot be analytically continued to $t>0$
($\rho>\rho_{\rm fp}$).  This result is in agreement with similar
derivations, e.g., in~\cite{Binder}.

It follows immediately from the above statement that the radius of
convergence of the virial expansion
\begin{equation}
  Z(\rho) = \sum_{n=1}^{\infty} B_n \rho^{n-1}
\end{equation}
is less than or at most equal to $\rho_{\rm fp}$.  In addition, approximation
$$
  B_n/\mathcal V^{n-1} = {\rm const}_0 + {\rm const}_1 n +{\rm const}_2 n^2,
$$
$n>n_0$, leading to the Carnahan-Starling and other popular equations
of state \cite{EOSs} of the form ${\rm polynomial}(y)/(1-y)^3$, is
not valid for sufficiently large $n$ and therefore such equations
cannot describe accurately a vicinity of the freezing point.
In the above formulas, $y=\mathcal V\rho$ is the packing fraction and
$\mathcal V$ is the sphere volume.

\subsection{Beyond stability}

Integral (\ref{ZanomG}) diverges in the metastable region
$\rho>\rho_{\rm fp}$ ($t>0$), but it can be \emph{approximately}
evaluated if the upper bound is identified with the maximum $r^*$ of
$G(r)$.  It means that configurations with droplets larger than the
critical size $r^*$ are omitted.  The error of this approximation is
estimated by (\ref{prec}).  In other words, the nonanalyticity at
$\rho=\rho_{\rm fp}$ does not allow extrapolation of $Z(\rho)$ to
$\rho>\rho_{\rm fp}$ with precision higher than given by (\ref{prec}).

\begin{figure}
\begin{center}
\includegraphics[scale=0.5]{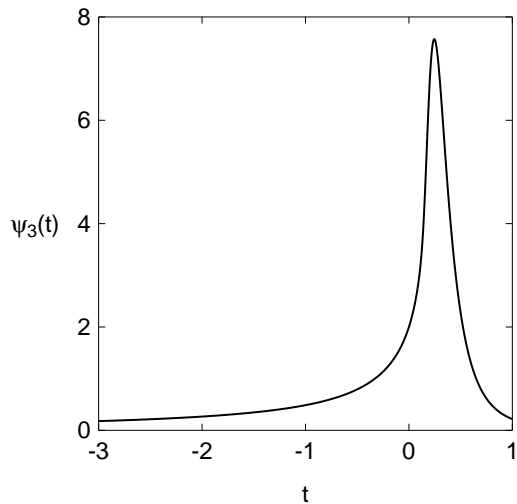}
\caption{Function $\psi_3(t)$, Eq.\ (\ref{psin}).}
\label{fig-psin}
\end{center}
\end{figure}

After normalization we get Eq.~(\ref{Zanom}) with the following extension of (\ref{psin0}),
\begin{equation}
\psi_n(t) = \int_{0}^{\mbox{\scriptsize$\cases{+\infty & for
      $t\le0$\cr 1/t  & for $t>0$}$}} x^n
 \exp\left({x^3\over3}t-{x^2\over2}\right){\rm d} x .
\label{psin}
\end{equation}
This function for $n=3$ is drawn in Fig.~\ref{fig-psin}.  The maximum,
appearing also on higher derivatives, matches the third hint of the
Introduction.

\subsection{Nonanalytical equation of state}

The equation of state should sew together both the low-density region
 and the anomalous part valid near the freezing
point.  We tried a simple sum of a polynomial in $\rho$ up to order
$\rho^{m-1}$ and the anomalous term,
\begin{equation}
  Z_m(\rho) = \sum_{i=1}^m C_i \rho^{i-1} + \beta
  \psi_3(\alpha(\rho-\rho_{\rm fp})) ,
\label{EOS}
\end{equation}
where coefficients $C_i$, $i=1,\dots,m$, are determined so that the
virial coefficients \cite{Labik,Clisby} up to $B_m$ are reproduced.
Two adjustable parameters $\alpha$ and $\beta$ were fitted to MD
data \cite{Kolafa} up to density $\rho_{\rm max}$.   
The value of $y_{\rm
  fp}=\pi\rho_{\rm fp}/6=0.494$ was taken from~\cite{freeze}. 
The standard deviation $\sigma$ of the fit
(the objective function) was 
\begin{equation}
  \sigma^2 = {1\over N_{\rm data}-2} \sum_{i=1}^{N_{\rm data}} \left[Z_i-Z_m(\rho_i)\over\sigma_i\right]^2,
\label{objf}
\end{equation}
where $\sigma_i$ is the standard error of molecular dynamics (MD)
datum $Z_i$, see below.

\begin{table}
\caption{MD data on the compressibility factor $Z$ at deeply
  metastable states.  The error bars are estimated standard deviations
  in units of the last significant digit, for $\rho\ge1.04$ combined
  with estimated metastable uncertainty.}
\label{tab-MDdata}
\begin{tabular}{cc}
\hline
$\rho$ & $Z$ \\\hline
1.02 & 16.55873(30) \\
1.03 & 17.2007(20) \\
1.04 & 17.847(20) \\
1.05 & 18.59(4) \\ 
1.06 & 19.41(8) \\
1.07 & 20.32(10) \\
1.08 & 21.30(15) \\
1.09 & 22.15(30) \\
\hline\end{tabular}
\end{table}

\section{Molecular dynamics data}

The values of the compressibility factors at reduced densities
$\rho\le1.02$ are taken from~\cite{Kolafa} where also
details of the molecular dynamics algorithm and finite-size
corrections are explained.  The $\rho=1.03$ value was recalculated and
higher densities newly calculated.  We combine runs of $N=1000$,
$2000$, $4000$, and $13500$ particles, but at lower densities we
disregard $N=1000$ results because of too large finite-size errors
(according to Eq.~(6) in~\cite{Kolafa}, the influence of periodic
boundary conditions is about 0.02 in $Z$ for the $\rho=1.04$, $N=1000$
system).

The time in MD calculations can be expressed either as real time $t$
in reduced time units (mass and sphere diameter are unity) or the
number of collisions $N_{\rm col}$ in the system.  Without finite size
corrections they are related by \cite{HSMD}
$$
  t = {N_{\rm col} \over N (3/\pi)^{1/2} (Z-1)} .
$$

Two algorithms for obtaining the initial configuration were used.  In
the first one \cite{Kolafa} a periodic cube with 500 spheres, obtained
by Monte Carlo simulation while shrinking to the desired density, was
replicated.  The second algorithm was based on soft spheres simulated
again by Monte Carlo method from random start with decreasing
temperature until all overlaps disappeared.  Then during a period of
60--100 time units (about 1000 collisions per particle) the system was
``partially equilibrated'', i.e., a plateau on the $Z(t)$ convergence
profile was reached.  The second algorithm gave better results because
in some cases the first one lead to a partly crystallized cube.

The metastable MD results need comment because the principle
inaccuracy limit is reached and data interpretation is not
straightforward.

At $\rho=1.02$ the system stays at the metastable state for
$2(1)\times10^{10}$ collisions and then it
``suddenly'' freezes.  The metastable chunks with approximately
constant values of $Z$ are long enough to be unambiguously extracted
and the main source of error is still the statistical uncertainty of the
data; the accuracy could be increased by a longer simulation.

The averaged freezing time for $\rho=1.03$ is $3.3(7)\times 10^8$
collisions ($t=1500$) for $N=13500$ and consistently $3.8(9)\times
10^8$ ($t=5000$) for $N=4000$ in agreement with the mechanism of
spontaneous homogeneous nucleation---the larger system, the higher
probability of nucleation.  However, the $N=1000$ system freezes
unexpectedly faster, in $6(2)\times 10^8$ collisions ($t=1250$), which
we attribute to the influence of the periodic boundary conditions.
The chunks of data corresponding to the metastable fluid are clearly
visible, but the points when a partly equilibrated metastable state
starts and where the first nucleus of the solid phase appears are not
exactly defined.  Our analysis of the data using a plot therefore
contains a subjective factor \footnote{Using subjective criteria in
data analysis may affect the results by unconscious bias.  One way to
avoid this bias would be a blind analysis by an independent researcher
similar to a blind experiment in drug testing.}  leading to scattering
of the $Z$ values by not more than 0.001 if the same procedure is
repeated, exemplifying the principle metastable uncertainty.  The
formally calculated error 0.0014 was therefore increased to 0.002.

For $\rho>1.04$ and partly also for $\rho=1.04$, $N>4000$, it becomes
difficult to determine the metastable parts as well as the onset of
freezing.  There is no plateau on the $Z(t)$ curve but rather a
shoulder \cite{Speedy}.  Apparent freezing appears within time of
roughly $t=200$ (4000 collisions per particle).  The main source of
error in the estimated $Z$ is the metastable uncertainty which is
inevitably rather guessed than calculated.  Our guesses are more
pessimistic than those by~\cite{Speedy}; in addition, our
$Z$ data are systematically by about 0.001 lower.  Our 
$\rho<1.04$ data also match
recent results \cite{Sadus} except the $\rho=1.04$ point which is
lower, $Z=17.76(2)$; there are no data in range $\rho\in[1.05,1.09]$ in
\cite{Sadus}.

\section{Results and discussion}

In spite of an obscure nature of error estimates in Table~\ref{tab-MDdata}
we tried to fit them to formula (\ref{prec}), assuming uniform
relative inaccuracies of these errors.  The resulting estimate of the
freezing density, $\rho_{\rm fp}=0.966(12)$, is surprisingly close to
the correct value 0.943 \cite{freeze}.


\begin{table}
\caption{Parameters $\alpha$ and $\beta$ of the nonanalytical EOS,
  Eq.~(\ref{EOS}), its residual standard deviation $\sigma$,
  Eq.~(\ref{objf}), and the residual standard deviation $\sigma_{\rm an}$ of
  the expansion in $x=y/(1-y)$ with two adjustable parameters, in
  dependence on the maximum virial coefficient $B_m$ included.}
\label{tab-res}
\begin{tabular}{c@{~~~~}cccc@{~~~~}cccc}\hline
 &\multicolumn{4}{c}{$\rho_{\rm max}=0.95$} 
 &\multicolumn{4}{c}{$\rho_{\rm max}=1.09$} \\
$m$ & $\alpha$ & $\beta$ & $\sigma$& $\sigma_{\rm an}$& $\alpha$ & $\beta$ & $\sigma$& $\sigma_{\rm an}$\\\hline
3 & 0.6594 & 12.81 & 347 & 36  & 0.6341 & 13.22 & 381 & 85  \\
4 & 0.5453 & 17.20 & 106 & 17  & 0.5337 & 17.69 & 106 & 33  \\
5 & 0.4740 & 23.02 & 34  & 17  & 0.4735 & 23.06 & 28  & 53  \\
6 & 0.4239 & 30.76 & 11  & 4.5 & 0.4371 & 28.66 & 26  & 18  \\
7 & 0.3877 & 40.56 & 5.1 & 8.4 & 0.4212 & 32.31 & 31  & 25  \\
8 & 0.3602 & 52.86 & 5.5 & 11  & 0.4266 & 31.06 & 30  & 22  \\
9 & 0.3407 & 66.40 & 6.2 & 8.6 & 0.4619 & 23.10 & 28  & 23  \\
10& 0.3353 & 71.69 & 6.3 & 21  & 0.5473 & 11.93 & 23  & 47  \\\hline
\end{tabular}
\end{table}

Results for the nonanalytical EOS (\ref{EOS}) with $\rho_{\rm
max}=0.95$ (only stable data except the last only slightly
metastable point) and $\rho_{\rm max}=1.09$ (all available data) are
collected in Table~\ref{tab-res}.  It is seen that the fitting is stable
and the resulting values of $\alpha$ and $\beta$ are in a physically
sensible range, especially if we do not consider $m\le4$ (not enough
virials to describe low density) and $m\ge 9$ (``too stiff'' virial
part).  It is also seen that the equation is not able to fit the data
within their standard deviations---we would need more adjustable
parameters.  Consequently, using $\sigma_i$ in (\ref{objf}) may be
considered inappropriate.  Therefore we tried also $\sigma_1=1$ (uniform
absolute error) and $\sigma_i=Z_i$, (uniform relative error), but the
results were qualitatively the same.

A ``standard'' approach to the HS EOS \cite{EOSs} is a polynomial in
$x=y/(1-y)$.  The coefficients up to power $x^m$ can be determined
from the known virial coefficients while higher coefficients are
fitted to the MD data.  Column $\sigma_{\rm an}$ of
Table~\ref{tab-res} contains the residual standard deviation
(\ref{objf}) of this analytical function with two adjustable
parameters.  It is seen that both the analytical and nonanalytical
EOSs are for $m\ge5$ comparable.  However, the result of the
nonanalytical EOS are, for $m\ge7$ for $\rho_{\rm max}=0.95$ or
$m\ge5$ for $\rho_{\rm max}=1.09$, uniformly good (or bad) while
adding one virial coefficient to the $x$-expansion may worsen the
result considerably.

\begin{figure}
\begin{center}
\includegraphics[scale=0.5]{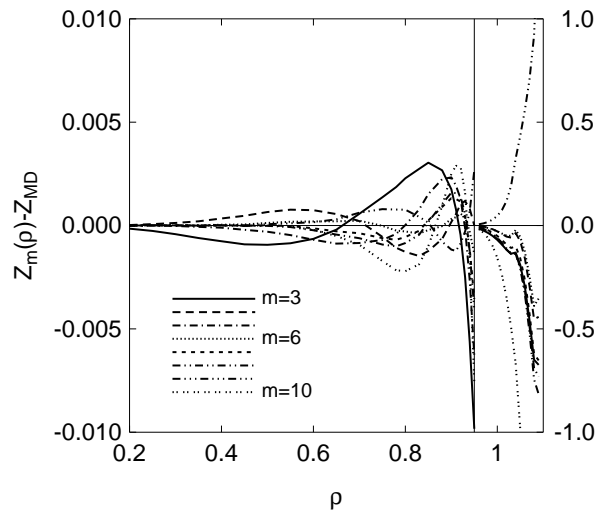}
\caption{Analytical equation of state (expansion in $x=y/(1-y)$):
  Deviations of the fit to $\rho_{\rm max}=0.95$ from MD data. 
  $B_m$ is the maximum virial included. 
  Note the change of the scale at $\rho=0.95$. }
\label{fig-x-exp}
\end{center}
\end{figure}

\begin{figure}
\begin{center}
\includegraphics[scale=0.5]{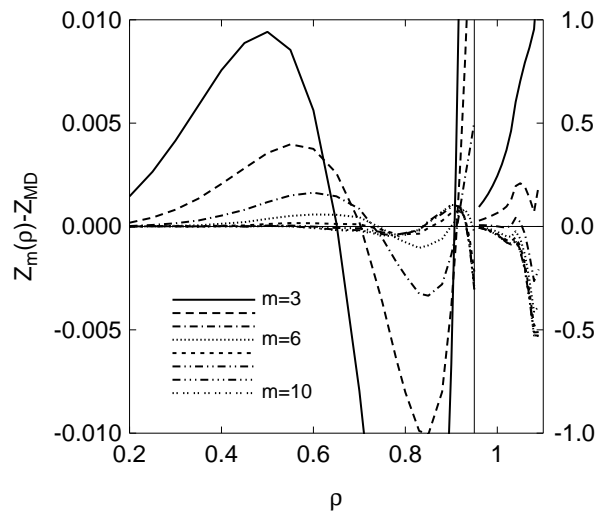}
\caption{Nonanalytical equation of state (\ref{EOS}).}
\label{fig-nonanal}
\end{center}
\end{figure}

Similar picture can be obtained from the ability of the EOSs fitted in
an (almost) stable range, $\rho_{\rm max}=0.95$, to extrapolate to
higher densities.  The analytical equation, Fig.~\ref{fig-x-exp}, is
better at low densities, but gives nonuniform extrapolation to the
deeply metastable region, especially if many virial coefficients are
taken into account.  In contrast, the nonanalytical equation,
Fig.~\ref{fig-nonanal}, though worse at low densities, gives (for
$m\ge5$) uniformly lower values.  The difference is only a few times
larger than the inherent metastable error and suggests that there is a
systematic inaccuracy in the model, in other words, the nonanalytical
term captures most but not all of the ``anomalous behavior''.

Both the analytical and nonanalytical equations exhibit a ``bump'' or
irregular behavior at $\rho=1.04$.  This is the density where
the well-defined plateaus on the $Z(t)$ curves cease to exist and data
interpretation contains a subjective factor.  From the microscopic
point of view the critical droplet size $r^*$ is comparable with
the molecule size and the nucleation model fails.

\section{Concluding remarks}

In spite of many approximations used (surface energy independent on
the droplet size, spherical droplets comparable to the atom size,
simple combination of the low-density expansion and high-density
region) the model of fluctuating droplets of solid state in fluid is
able to provide stable results on the equation of state near freezing
density and enables extrapolation to higher densities.

Unfortunately, integral (\ref{psin}) is probably too complicated for
practical application.

\section*{Acknowledgments}  
This work was supported by the The Ministry of
Education, Youth and Sports of the Czech Republic under the project
LC512 (Center for Biomolecules and Complex Molecular Systems).
I wish also thank Prof.\ Roman Koteck\'y (Charles University, Prague)
for helpful discussions about Ising systems.

\end{document}